\PassOptionsToPackage{unicode}{hyperref}
\PassOptionsToPackage{hyphens}{url}
\PassOptionsToPackage{dvipsnames,svgnames,x11names}{xcolor}
\documentclass[
  12pt]{article}

\usepackage{amsmath,amssymb,amsthm}
\usepackage{graphicx,psfrag,epsf}
\usepackage{enumerate}
\usepackage{enumitem}
\usepackage{natbib}
\usepackage{url} 
\usepackage{mathtools,amssymb,bm}
\usepackage{multicol}
\usepackage{multirow}
\usepackage{makecell}

\usepackage{iftex}
\ifPDFTeX
  \usepackage[T1]{fontenc}
  \usepackage[utf8]{inputenc}
  \usepackage{textcomp} 
\else 
  \usepackage{unicode-math}
  \defaultfontfeatures{Scale=MatchLowercase}
  \defaultfontfeatures[\rmfamily]{Ligatures=TeX,Scale=1}
\fi
\usepackage{lmodern}
\ifPDFTeX\else  
\fi
\IfFileExists{upquote.sty}{\usepackage{upquote}}{}
\IfFileExists{microtype.sty}{
  \usepackage[]{microtype}
  \UseMicrotypeSet[protrusion]{basicmath} 
}{}
\makeatletter
\@ifundefined{KOMAClassName}{
  \IfFileExists{parskip.sty}{%
    \usepackage{parskip}
  }{
    \setlength{\parindent}{0pt}
    \setlength{\parskip}{6pt plus 2pt minus 1pt}}
}{
  \KOMAoptions{parskip=half}}
\makeatother
\usepackage{xcolor}
\makeatletter
\ifx\paragraph\undefined\else
  \let\oldparagraph\paragraph
  \renewcommand{\paragraph}{
    \@ifstar
      \xxxParagraphStar
      \xxxParagraphNoStar
  }
  \newcommand{\xxxParagraphStar}[1]{\oldparagraph*{#1}\mbox{}}
  \newcommand{\xxxParagraphNoStar}[1]{\oldparagraph{#1}\mbox{}}
\fi
\ifx\subparagraph\undefined\else
  \let\oldsubparagraph\subparagraph
  \renewcommand{\subparagraph}{
    \@ifstar
      \xxxSubParagraphStar
      \xxxSubParagraphNoStar
  }
  \newcommand{\xxxSubParagraphStar}[1]{\oldsubparagraph*{#1}\mbox{}}
  \newcommand{\xxxSubParagraphNoStar}[1]{\oldsubparagraph{#1}\mbox{}}
\fi
\makeatother

\usepackage{longtable,booktabs,array}
\usepackage{calc} 
\usepackage{etoolbox}
\makeatletter
\patchcmd\longtable{\par}{\if@noskipsec\mbox{}\fi\par}{}{}
\makeatother
\IfFileExists{footnotehyper.sty}{\usepackage{footnotehyper}}{\usepackage{footnote}}
\makesavenoteenv{longtable}
\usepackage{graphicx}
\makeatletter
\def\maxwidth{\ifdim\Gin@nat@width>\linewidth\linewidth\else\Gin@nat@width\fi}
\def\maxheight{\ifdim\Gin@nat@height>\textheight\textheight\else\Gin@nat@height\fi}
\makeatother
\setkeys{Gin}{width=\maxwidth,height=\maxheight,keepaspectratio}
\makeatletter
\def\fps@figure{htbp}
\makeatother

\makeatletter
\@ifpackageloaded{caption}{}{\usepackage{caption}}
\AtBeginDocument{%
\ifdefined\contentsname
  \renewcommand*\contentsname{Table of contents}
\else
  \newcommand\contentsname{Table of contents}
\fi
\ifdefined\listfigurename
  \renewcommand*\listfigurename{List of Figures}
\else
  \newcommand\listfigurename{List of Figures}
\fi
\ifdefined\listtablename
  \renewcommand*\listtablename{List of Tables}
\else
  \newcommand\listtablename{List of Tables}
\fi
\ifdefined\figurename
  \renewcommand*\figurename{Figure}
\else
  \newcommand\figurename{Figure}
\fi
\ifdefined\tablename
  \renewcommand*\tablename{Table}
\else
  \newcommand\tablename{Table}
\fi
}
\@ifpackageloaded{float}{}{\usepackage{float}}
\floatstyle{ruled}
\@ifundefined{c@chapter}{\newfloat{codelisting}{h}{lop}}{\newfloat{codelisting}{h}{lop}[chapter]}
\floatname{codelisting}{Listing}

\makeatother
\makeatletter
\@ifpackageloaded{caption}{}{\usepackage{caption}}
\@ifpackageloaded{subcaption}{}{\usepackage{subcaption}}
\makeatother

\ifLuaTeX
  \usepackage{selnolig}  
\fi
\usepackage[]{natbib}
\usepackage{bookmark}

\IfFileExists{xurl.sty}{\usepackage{xurl}}{} 
\hypersetup{
  pdftitle={Mixed Frequency Stochastic Frontier Model: with application to the linkage of weather extremes and firm efficiency},
  pdfauthor={Erniel Barrios, Nur Syazwani Mazlan, Lim Foo Weng, Paolo Victor Redondo, Gian Karlo M. Torreno, Lee How Chinh},
  pdfkeywords={Additive models, backfitting algorithm, extreme weather events, production efficiency, semiparametric model, stochastic frontier model},
  colorlinks=true,
  linkcolor={blue},
  filecolor={Maroon},
  citecolor={Blue},
  urlcolor={Blue},
  pdfcreator={LaTeX via pandoc}}

\newcommand{\anon}{1}

\begin{document}

\def\spacingset#1{\renewcommand{\baselinestretch}%
{#1}\small\normalsize} \spacingset{1}


\if1\anon
{
  \title{\bf Mixed Frequency Stochastic Frontier Model: with application to the linkage of weather extremes and firm efficiency}
  \author{Erniel Barrios$^{\text{\textbf{a}}}$\footnote{Corresponding author: erniel.barrios@monash.edu}, Nur Syazwani Mazlan$^{\text{\textbf{a}}}$, Lim Foo Weng$^{\text{\textbf{b}}}$\\
  , Paolo Victor Redondo$^{\text{\textbf{c}}}$, Gian Karlo Torreno$^{\text{\textbf{a}}}$, and Lee How Chinh$^{\text{\textbf{a}}}$\\
  {\small $^{\text{\textbf{a}}}$School of Business, Monash University Malaysia, erniel.barrios@monash.edu}\\ 
  {\small $^{\text{\textbf{b}}}$School of Mathematical Sciences, Sunway University}\\ 
  {\small $^{\text{\textbf{c}}}$King Abdullah University of Science and Technology, paolovictor.redondo@kaust.edu.sa}}
  \maketitle
} \fi

\if0\anon
{
  \bigskip
  \bigskip
  \bigskip
  \begin{center}
    {\LARGE\bf Mixed Frequency Stochastic Frontier Model: with application to the linkage of weather extremes and firm efficiency}
\end{center}
  \medskip
} \fi

\bigskip
\begin{abstract}

This paper proposes a novel stochastic frontier framework that incorporates high-frequency determinants of inefficiency into models where production indicators are observed at lower frequencies. The approach addresses the persistent issue of positive skewness in stochastic frontier analysis by specifying inefficiency through a logistic function, thereby avoiding the need for restrictive and often complex distributional assumptions for the inefficiency term. High-frequency covariates are incorporated into the inefficiency equation using a flexible nonparametric function, enabling the model to capture heterogeneous and time-varying influences on firm performance. Estimation is conducted using a hybrid backfitting algorithm rather than conventional maximum likelihood estimation, mitigating convergence problems and reducing sensitivity to distributional misspecification. The proposed methodology is applied to a panel of electric cooperatives in the Philippines to examine the relationship between firm efficiency and extreme weather events. The results demonstrate how climate-related shocks can be linked to operational inefficiency at the firm level, providing a practical framework for evaluating micro-level sustainability and resilience challenges associated with climate change. This approach broadens the applicability of stochastic frontier analysis to contexts characterized by mixed-frequency data and complex environmental determinants of efficiency.

\end{abstract}

\noindent%
\vfill

\noindent%
\vfill

\newpage
\spacingset{1.8} 

\section{Introduction}\label{chap:introduction}

The Philippines’ power sector in the 1990s faced severe challenges confronting the duality of electricity generation and distribution. Republic Act No. 9136, also known as the Electric Power Industry Reform Act of 2001 (EPIRA), was aimed to lay the groundwork for an open market that stimulates competition, facilitating the entry of more private investments into the sector. The legislative agenda was to ensure reliable and affordable electricity in the country for both the urban centers and rural areas. Two regulatory agencies were formed in the process, including the Energy Regulatory Commission (ERC) and the Power Sector Assets and Liabilities Management Corporation (PSALM).

The implementation strategy of EPIRA to stir competition in the power market includes the creation of the Wholesale Electricity Spot Market (WESM) featuring competitive market forces that lead to the generation tariffs and a more transparent cost structure \citep{brucal2018}. The lower cost and accessibility feature resulting from the improvement in generation, transmission and distribution of electricity drive a wider reach of electricity, especially in mostly isolated rural areas. This has not only expanded the reach of electricity access but the industry is also leading more towards fiscal independence.

The landmark law and other legislative provisions led the electricity industry in the Philippines to go through various reforms with the general theme of privatization \citep{sharma2004}. In the context of scale efficiency, certain operations are still kept as public, others are privatized, hoping that competition will drive efficiency, which could be loosely interpreted to be parallel or interchangeable. As more privatization is seen in the sector, regulation is becoming more important than before. In the Philippines, \citet{pacudan2002} noted that technical inefficiency, which is reflected in terms of system loss, is primarily driven by scale inefficiency. This provides a stimulus for various policies that can result in overarching benefits to the whole economy while the energy sector demonstrates production efficiency. Policies and other regulatory instruments in the Philippines have favored aiming for production efficiency, especially in electricity distribution.

Privatization of electricity distribution in the Philippines was founded on the idea of cooperativism, i.e., consumers organize into a micro-enterprise and manage the operations of electricity distribution. This distribution set-up has been instrumental for the sustainable and reliable electricity in rural areas, noted by \citet{escresa2024}. While this approach has been effective in developed countries like the USA, it has been challenging for developing countries, especially those with weak institutions or weak competition. This loophole in the strategy necessitates the review of regulatory policies. Another law (RA No. 10531) mandates the National Electrification Administration (NEA) to ensure the viability of electric cooperatives (EC) in retail competition and open access \citep{nea2025}. NEA is further mandated to provide support and interventions, including capacitating ECs with financial and technical knowledge to enhance operational efficiency. As of the first quarter of 2025, there are a total of 121 ECs with varying financial standing throughout the country.

Typhoons and rainfall in the Philippines had worsened in the recent past. There are around 20 typhoons in a year, and around 8 or 9 pass through the Philippine Area of Responsibility (PAR). Peaks occur around July--October and are mostly destructive \citep{santos2021}. Geographical features, as well as the physical environments, make the country susceptible to natural disasters including landslides, flooding, and destructive strong winds. \citet{racoma2022} noted that while typhoons are instrumental in precipitation needed for agriculture, hazards like flooding, landslides, and strong winds are observed consequences. However, \citet{olaguera2022} noted that between the 1950s and 2000s, there was an increasing trend in rainfall and the number of days with rainfall above 50 mm, an increasing trend of maximum 5-day rainfall (extreme), and decreasing trends in the length of dry spells (between July and September).

The NEA is also mandated to monitor financial indicators that characterize the operational status of ECs with the intention of ensuring operational efficiency. However, as the Philippines is along the pathway of typhoons, and coupled with monsoon rains that occur around the same time as the later part of the typhoon season, operations of ECs have become vulnerable due to the exposure of the distribution infrastructure to these weather extremes. As an example, poles for distribution lines can be easily toppled by strong winds or buried by landslides in hilly or mountainous areas. Even if the EC is financially viable, operational efficiency is challenged by the impact of extreme weather events.

The goal of this paper is to assess the impact of climate change (indexed by extreme windspeed and extreme rainfall) on firm efficiency (focusing on EC and the distribution of electricity). The empirical linkage between firm efficiency and climate change will be built along the framework of stochastic frontier model (SFM). The nature of the data (panel) and the circumstances associated with the data generating process leads to two modeling issues: mixed frequency time series and skewness in the inefficiency error component.

Firm (EC) characteristics are mostly related to financial ratios or operational indicators that are aggregated annually. Climate change will be characterized by extreme weather events (maximum rainfall and maximum windspeed during typhoon) exhibiting large variability over the months when tropical cyclones pass the Philippines’ area of responsibility, and very low or zero values during the dry season. The original proposal to resolve the varying frequency time series problem was to aggregate high frequency (monthly) data to be aligned with the low frequency (annual) data. This however, is prone to information loss as there is significant variability of windspeed and rainfall over the months within a year, windspeed can go as high as over 300 KPH with strong typhoons to 0 during dry months. \citet{ghysels2007} presented mixed data sampling regression (MIDAS) based on a distributed lag model to avoid aggregation in mixed frequency time series data. There are some enhancements to MIDAS, e.g., \citet{ghysels2016} introduced mixed frequency vector autoregression for a more parsimonious parametrization, and \citet{gotz2014} embedded MIDAS into an error correction model. There are many other solutions to the issues with mixed frequency data, e.g., \citet{seong2020} introduced a smoothing method at low frequency to impute data at high frequencies, and \citet{schorfheide2015} introduced the state space and Bayesian framework into vector autoregressions. A nonparametric function was used to incorporate high frequency predictors in volatility models \citep{benito2025}, in spatiotemporal models \citep{malabanan2022}, and to account for structural change \citep{glova2025}. Due to sparsity in extreme rainfall and windspeed, these will be introduced in the inefficiency equation of SFM as nonparametric functions.

Due to issues pertaining to regulation, the ``positive skewness'' problem common in SFM is expected to arise. The literature of SFM resolves this problem through the use of a half-normal (truncated) distribution (and other extensions) for the inefficiency error component. The complexity of the likelihood function, however, has always been a threat to the convergence of the maximum likelihood function estimation. These issues will be resolved in this paper by invoking a logistic function for the inefficiency equation and estimation of parameters and nonparametric functions through hybrid methods in the backfitting algorithm framework.

The paper is organized as follows: We present in Section 2 methods of measuring firm efficiency and the associated challenges in using stochastic frontier models. The methodology is discussed in Section 3, where the data and indicators are thoroughly described, the model is formulated, and estimation procedure is characterized. Results are presented and discussed in Section 4, including issues on convergence, comparison with SFM using aggregated extreme rainfall and windspeed (annual), determinants of inefficiency, and the quantification of technical efficiency (TE) over time and over space (regions). Some concluding notes are presented in Section 5.

\section{Efficiency of Firms}\label{chap:model}

The economic literature specifies that production is a result from activities that uses labor and capital as main inputs. Given a production technology, increasing labor and capital inputs leads to increasing production. There are however, some constraints on the increase in production due to the carrying capacity of the production system, which leads to the specification of the exponential-based production function to instill these production limits. Some intervening and moderating factors are included in the model, but still, the predictive ability of the model is insufficient since the error of the production function is not just pure error but can also include the distance of production system from the production frontier (inefficiency of production units).

\subsection{Stochastic Frontier Models}

Analysis of efficiency relies on the specification and estimation of production functions that describe the relation between inputs and outputs in the production system. Firm-level analyses derive such estimates from a simple Cobb--Douglas production function. It is important that modelers are careful in model specification, such as functional form and error distribution, to avoid biased estimates or misleading inferences \citep{zellner1966, griffin1987}. A context-dependent strategy, trading off theoretical consistency against empirical fit, is crucial to provide robust production function analysis.

The stochastic frontier models (SFM) were introduced around 1977 by \citet{aigner1977} and \citet{meeusen1977}. \citet{kumbhakar2003} noted that the topic has grown so immensely that SFMs are now a part of efficiency analysis since they enable the decomposition of the error term into inefficiency and pure error. SFMs has been used in diverse applications, e.g., Indonesian pharmaceutical industry, Norwegian grain farming \citep{kumbhakar2014}, Turkish agriculture \citep{dudu2015}, impact of credit constraints on hybrid maize farmers in Pakistan \citep{ali2019}, influence of corruption control on efficiency across 102 countries \citep{kutlu2023}, application to soil conversion in El Salvador \citep{centorrino2024}, and many others.

\subsection{Challenges in Stochastic Frontier Models}

Despite their advantages and utility in various applications, SFMs face several challenges, including convergence problems and the need for alternative parameterizations. While SFMs were first presented in a fully parametric form, an appropriate functional form and specification of the distribution of noise remain a big challenge in the estimation. Thus, deciding on one type of production function over another or in identifying which exogenous variables might impact inefficiency remains a complex task \citep{amsler2009}.

Efficient estimates are hinged on model specification. \citet{kumbhakar2014} emphasized the need to correct for heteroskedasticity and firm-specific effects, as these significantly influence the estimates. Moreover, \citet{campos2022} addressed convergence issues by introducing new parametrizations of the skew normal distribution, creating a much more stable framework for estimating efficiency.

Recent advancements in SFMs have progressed to achieve significant improvements in solving some of the issues, like the ``wrong skewness problem'', for example, \citet{almanidis2012}, \citet{hafner2018}, \citet{papadopoulos2024}, and \citet{haschka2025}. This problem occurs when the estimated skewness of the error term does not match the theoretical expected skewness. Some of the proposed solutions are correct, or adjust the distribution of the noise component, and enhance flexible functional forms. \citet{ferrara2017} developed a semiparametric method for the production function that was subsequently extended by \citet{forchini2023}. This new formulation provided greater flexibility, relaxing the restrictive assumptions typically placed on the production function.

\subsection{Recent Advances in Stochastic Frontier Models}

Spatial dependencies are now an indispensable tool in the analysis of efficiency, especially in cases where regional spillovers and interdependencies are important. Recent work in the field has concentrated on minimizing spatial dependence, heteroscedasticity, and endogeneity, and on supplying strong frameworks to underpin policy-relevant applications. A common theme in many of these studies is the utilization of latent spatial structures to account for unobserved geographical heterogeneity, e.g., \citet{schmidt2009} and \citet{gude2018}. Studies also focused on the development of models to address asymmetric efficiency spillovers along with endogeneity, see for example, \citet{glass2016} and \citet{kutlu2020}.

Methodological improvements to address spatial dependence and technical inefficiency are presented in \citet{tsukamoto2019}, \citet{battese1995}, \citet{degraaff2020}, and \citet{galli2023}, among others. The development of spatial SFMs has emphasized latent spatial patterns, heterogeneous effects, interconnectedness, and flexible specifications. These have refined the efficiency estimates with greater accuracy and offered useful information to policymakers in addressing regional imbalances and promoting economic cohesion.

Bayesian inference procedures also contributed to SFMs to address issues like temporal behavior of technical efficiency, spatial dependence, and unobserved heterogeneity, see for example, \citet{tsionas2006}, \citet{schmidt2009}, \citet{carvalho2018}, and \citet{wanke2020}, among others. \citet{campos2022} pointed out the practical advantages of Bayesian inference procedures in the midst of regulation, highlighting the complementary application of SFM and Data Envelopment Analysis (DEA), more important observation is that although SFM tends to be more robust in the presence of outliers, it experiences convergence-related issues that can be resolved in Bayesian framework.

One of the shortcomings of the conventional SFMs is the assumption of time-invariant inefficiency, which combines inefficiency with time-invariant cross-unit heterogeneity. \citet{greene2005} extended the framework of the stochastic frontier by allowing fixed and random effects specifications. \citet{kumbhakar2014} specified a model that decomposes firm effects into persistent (time-invariant) and residual (time-varying) technical inefficiency, emphasizing the need for proper handling of inefficiency and heterogeneity in empirical practice.

Recent developments have extended the range of SFMs to allow for more elaborate specifications of inefficiency and assumptions of distributions. \citet{elmehdi2025} addressed the issue of right-skewed residuals in SFMs with a panel data model in an extended-half-normal distribution of the inefficiency term, useful in solving model misspecification issues and detecting significant inefficiency. In addition, \citet{haschka2025} also questioned the universal supposition that inefficiency terms are always positively skewed and estimated a model where both positive and negative skewness are present. Another recent work by \citet{cai2024} developed a nonparametric estimator for SFMs allowing a discrete mass at zero inefficiency, enabling detection of fully efficient firms under a unimodality-at-zero assumption.

\section{Methodology}\label{chap:method}

The power crisis in the Philippines in the 1990s led to the review of certain policies and updating of the legislative framework governing the sector. Major structural transformations pursued were the privatization of electricity distribution and lesser regulation for electricity generation, which are intended to encourage investors and open the sector from monopoly, as this hinders growth and efficiency of the sector. Privatization and less regulation in Turkey, according to \citet{bagdadioglu1996}, resulted in better technical and scale efficiency of the energy sector.

Electricity distribution has been an important theme of research in the energy sector. Using data envelopment analysis (DEA), \citet{chen2002} investigates various regulatory policies in the distribution of electricity that leads to efficiency of distribution utilities. DEA was also used by \citet{mullarkey2015} to assess the impact of reorganization of the distribution network for electricity that leads to greater efficiency. However, \citet{kumbhakar2015} observed that scale economies have the potential to drive efficiency in electricity distribution.

The effect of extreme weather on electricity distribution has been the recent theme of researchers and engineers in the energy sector. For instance, \citet{goncalves2024} characterized the major impact of extreme weather events on energy systems. \citet{nyangon2024} noted the increased risk of power systems from climate change and that utility companies should include investments in assets that are resilient to these extreme weather episodes. \citet{hawker2024} and \citet{shen2024} added that this investment is needed not only in rural but also in urban areas.

There is a long history of Stochastic Frontier Models (SFM) since its introduction by \citet{aigner1977} and \citet{meeusen1977}, but considerable issues on the structural form and distributional assumptions remain. Recent literature in SFM pointed out the presence of the ``wrong skewness'' problem, i.e., the distribution of inefficiency is positively skewed. The model views firms as more efficient than they really are. \citet{papadopoulos2024} provided an illustration of an existing economic structure that exhibits positive skewness of the inefficiency error component. This is the case when firms are heavily regulated or when stringent barriers are present for the entrance of inefficient firms. Given the half-normal distribution for the inefficiency error, maximum likelihood estimation (MLE) will estimate inefficiency as zero \citep{kumbhakar2013} and the resulting information matrix is non-singular, leading often to non-convergence of MLE. \citet{hafner2018} generalized the distribution of inefficiency error to produce non-degenerate estimates of inefficiency even with positive (wrong) skewness. \citet{haschka2025} used a closed skew normal distribution, while \citet{horrace2024} used quantile estimation to resolve the problem.

In this paper, the functional form of inefficiency error function is postulated instead of fine-tuning the distribution of decomposed error components in the model. Specifically, a logistic function for the inefficiency components was used to induce non-negativity of the inefficiency error. Furthermore, an additive nonparametric function was added to the logistic function to account for extreme weather indicators in the inefficiency function. Model additivity facilitate estimation through the hybrid methods in the backfitting algorithm framework, with guaranteed convergence \citep{ansley1994} and other optimal and asymptotic properties \citep{mammen1999, opsomer2000}. The backfitting algorithm further mitigates convergence issues usually encountered in MLE (assuming, say, the half-normal distribution of the inefficiency term).

\subsection{A Spatio-temporal Data}

The data on 114 electric cooperatives (EC) from 2010 to 2022 were obtained from the National Electrification Administration (https://www.nea.gov.ph/ao39/). Financial profile (various financial ratios), including indicators of inputs, operations and outputs is reported annually by the cooperatives. The country is composed of 17 regions, but region 13 (National Capital) was excluded since only one cooperative is present (no competition) in the region that is highly urbanized. Several cooperatives operate within the scope of a region, ranging from as few as 3 to as many as 14. Denote $y_{it}$, $i=1,2,\ldots,n$ (number of electric cooperatives), $t=1,\ldots,T$ (number of years of EC profile) as the production/output indicators. The gross operating revenue (GOR) serves as the production indicator for the ECs. Since the ECs exhibit large variability in terms of the size of operations, heterogeneity of GOR is effectively controlled by capital indicators which captures the size of the operation of the EC. The labor indicator is the actual number of employees (EMPL) denoted by $L_{it}$. Several capital indicators are considered, including: $C_1=\text{peak load (PEAK)}$; $C_2=\text{number of consumers (CONS)}$; $C_3=\text{remittance to PSALM (PSALM)}$, and; $C_4=\text{reinvestment fund (REINV)}$. PEAK (indicating the capacity of the distribution system), CONS, PSALM, and REINV represents various investments/assets of the cooperatives. Determinants of efficiency/inefficiency considered are $Z_1=\text{collection efficiency (CE)}$; $Z_2=\text{system loss (SL)}$, and; $Z_3=\text{interest expenses (IE)}$. Note that while the GOR has been increasing over the years, it peaked between 2018 and 2019. Meanwhile, the efficiency/inefficiency indicators are fluctuating over similar period, see Table 1 for details. GOR is strongly correlated with REINV (0.8804) and PEAK (0.7452) and exhibited positive correlations with PSALM and CONS. On the other hand, GOR is negatively correlated with efficiency/inefficiency indicators except to IE.

\begin{table}[!t]
\centering
\caption{Median for Output, Input, and Efficiency/Inefficiency Indicators (By Year)}
\label{tab:median_indicators}
\begin{tabular}{ccccccccc}
\hline
\textbf{Year} & \textbf{GOR} & \textbf{PEAK} & \textbf{CONS} & \textbf{PSALM} & \textbf{REINV} & \textbf{CE} & \textbf{SL} & \textbf{IE} \\
\hline
2010 & 596,796  & 20,773 & 60,470  & 6,562  & 21,162 & 96.41 & 12.733 & 3,711 \\
2011 & 687,514  & 21,410 & 62,643  & 4,662  & 26,666 & 96.02 & 12.168 & 3,850 \\
2012 & 708,910  & 22,566 & 66,496  & 10,305 & 27,726 & 96.33 & 12.131 & 4,471 \\
2013 & 828,905  & 25,240 & 68,148  & 22,084 & 29,738 & 95.32 & 12.065 & 4,644 \\
2014 & 864,500  & 24,628 & 73,969  & 29,690 & 31,572 & 95.80 & 11.952 & 4,827 \\
2015 & 961,212  & 25,270 & 77,531  & 38,341 & 34,337 & 97.33 & 11.601 & 4,659 \\
2016 & 1,080,904 & 27,236 & 80,700  & 46,023 & 37,847 & 97.89 & 11.264 & 5,063 \\
2017 & 1,271,305 & 28,411 & 84,808  & 62,030 & 42,253 & 98.13 & 10.818 & 4,459 \\
2018 & 1,512,538 & 22,438 & 31,314  & 81,957 & 47,048 & 91.76 & 14.000 & 6,067 \\
2019 & 1,620,753 & 23,440 & 31,784  & 66,173 & 49,535 & 90.54 & 13.100 & 6,578 \\
2020 & 1,477,255 & 30,383 & 100,389 & 47,318 & 51,388 & 95.47 & 9.946  & 5,890 \\
2021 & 1,802,667 & 31,702 & 104,939 & 44,973 & 56,662 & 96.62 & 9.918  & 6,560 \\
2022 & 2,368,461 & 32,580 & 98,070  & 46,709 & 47,902 & 97.70 & 9.638  & 6,678 \\
\hline
\end{tabular}
\end{table}

Monthly Rainfall from different monitoring stations (2010--2022), and windspeed of typhoons that enter the Philippine Area of Responsibility (PAR) were obtained from the Philippine Atmospheric, Geophysical and Astronomical Services Administration (\url{https://bagong.pagasa.dost.gov.ph/climate/climate-data}). The data was parsed to generate the following weather indicators:

\begin{itemize}
\item $W_{t_j}$ -- Maximum Sustained Wind for month $j=1,2,\ldots,12$ in year $t$ (0 if no typhoon in a month). Typhoons passing through the PAR are not localized over the regions; thus, similar values for $W_{t_j}$ are assumed for all regions.

\item $R_{t_j}$ -- Maximum Rainfall for month $j=1,2,\ldots,12$ in year $t$ across all sampling stations in the same region. Thus, different values of $R_{t_j}$ are recorded across different regions.
\end{itemize}

Maximum rainfall exhibits patchiness in the distribution across the months over different regions (see Figure 1). Several typhoons could pass through PAR in a month, and to emphasize extreme events in a month, e.g., several typhoons occurring in a month, the maximum sustained winds for all typhoons in a month were added. This is the reason why some entries in Table 2 are over 800, possibly because many typhoons could pass PAR in a given month. Together with maximum sustained winds (mostly super typhoon levels) from Table 2, this can potentially cause vulnerability to the efficiency of production by the electric cooperative temporally and over space (regions).

\begin{table}[!t]
\centering
\caption{Summary Statistics for Extreme Weather Indicators by Year}
\label{tab:weather_summary}
\begin{tabular}{cccccc}
\hline
\multirow{2}{*}{\textbf{Year}} &
\multirow{2}{*}{\textbf{GOR}} &
\multicolumn{2}{c}{\textbf{Maximum Sustained Wind}} &
\multicolumn{2}{c}{\textbf{Maximum Rainfall}} \\
\cline{3-6}
& & \textbf{Median} & \textbf{Mean} & \textbf{Median} & \textbf{Mean} \\
\hline
2010 & 596,796   & 375 & 375 & 787.4 & 749.8 \\
2011 & 687,514   & 425 & 425 & 978.6 & 901.2 \\
2012 & 708,910   & 555 & 555 & 849.2 & 927.0 \\
2013 & 828,905   & 800 & 800 & 745.7 & 824.0 \\
2014 & 864,500   & 550 & 550 & 888.6 & 820.1 \\
2015 & 961,212   & 400 & 400 & 615.0 & 662.6 \\
2016 & 1,080,904 & 655 & 655 & 764.9 & 798.1 \\
2017 & 1,271,305 & 385 & 385 & 751.0 & 802.7 \\
2018 & 1,512,538 & 650 & 650 & 885.4 & 871.0 \\
2019 & 1,620,753 & 535 & 535 & 738.6 & 756.8 \\
2020 & 1,477,255 & 565 & 565 & 549.5 & 640.1 \\
2021 & 1,802,667 & 335 & 335 & 880.0 & 881.4 \\
2022 & 2,368,461 & 895 & 895 & 756.3 & 696.8 \\
\hline
\end{tabular}
\end{table}

\begin{figure}[!t]
    \centering
    \includegraphics[width=0.5\linewidth]{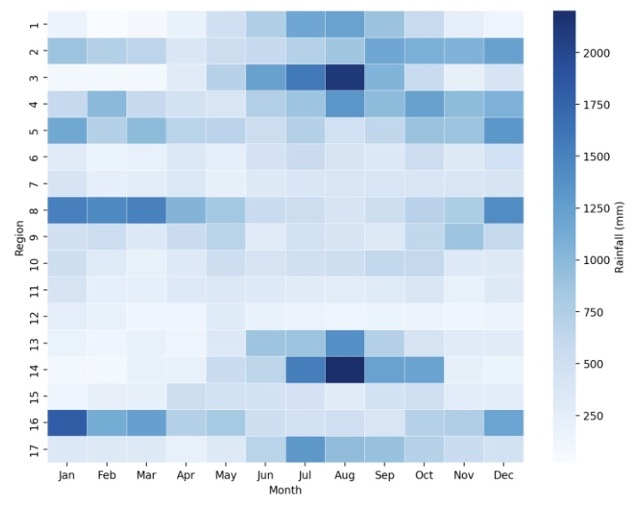}
    \caption{Maximum Monthly Rainfall by Region (2010-2022)}
    \label{fig:max_rainfall}
\end{figure}

\subsection{Stochastic Frontier Model with Varying Frequencies}

Productivity (profitability) of firms, in this case, EC, is not just a function of labor and capital, but also includes some moderating and intervening factors. Certain indicators like collection efficiency ($Z_1$), system loss ($Z_2$), and interest expenses ($Z_3$) can mediate production of ECs, but instead of integrating these into the production function, we specify them as determinants of efficiency/inefficiency of firms, serving as hindrance or facilitating factors towards the achievement of the production frontier.

The spatial-temporal stochastic frontier model \citep{barrios2021} is simplified to a panel data model with spatial externalities subsumed into the inefficiency equation through extreme weather indicators that will have a spanning effect across the areas where the ECs are located. Suppose the output of firm $i$ at time $t$ ($y_{it}$) is linked to labor indicators ($L_{it}$) and capital indicators ($C_{1it}, C_{2it}, C_{3it}, C_{4it}$) through the production function:
\begin{equation}
\ln y_{it} = \alpha_0 + \alpha_1 \ln L_{it} + \sum^4_{j=1}\beta_j \ln C_{jit} + v_{it} - u_{it}, \qquad
\text{where} \qquad v_{it}=\rho v_{i(t-1)}+\psi_{it}.
\end{equation}

The parameters $\alpha_1$ and $\beta_j$'s represents the elasticity to labor and capital, while $\rho$ captures temporal dependencies in the data. The pure error $\psi_{it} \sim N(0,\sigma_{\psi}^{2})$ represents the random shocks to the production output for firm/EC $i$ at time $t$. Spatial externalities captured by extreme weather conditions are captured in the panel data and accounted for in the model through the inefficiency equation given in Equation (2). Note that ECs are independent entities and while some are located in the same region, they are related (spatially) only in terms of similar vulnerabilities to extreme weather conditions in the region.

Suppose that (in)efficiency-enhancing indicators are measured for all $i$ and $t$ ($Z_{1it}, Z_{2it}, Z_{3it}$). This will also capture heterogeneity of the firms in the panel and facilitate pooling of the data together across all firms (EC) over time.

The variables $y_{it}$, $L_{it}$, $C_{it}$, and $Z_{it}$ are measured at the same frequency ($t$), i.e., annually for each firm $i$ from the data described in Section 3.1. Extreme weather events vary over the months within the year. In the Philippines, extremely dry weather typically occurs from March to May, while tropical typhoons usually occur from July to October. Extreme drought is captured by maximum rainfall for the month equal to zero, while strong typhoons are measured by maximum wind speed for the month, zero in months without typhoons. Aggregation of monthly data to annual levels to match low frequency indicators could mask the possible influence of monthly fluctuations on the efficiency of firms. Both $W_{t_j}$ and $R_{t_j}$ are sparse and mostly zero during dry months and when there is no typhoon. As an illustration, when rainfall is too much, this can loosen the soil cover and could result in landslides, causing the risk of transmission lines falling down. On the other hand, when the typhoon is so strong, winds could easily topple transmission lines which can disrupt operation of the EC, and restoration will add to the operational cost of the EC. Thus, aggregation of $W_{t_j}$ and $R_{t_j}$ to annual levels will neutralize extreme windspeed during typhoons and extreme rainfall during the monsoon season. To induce non-negativity to the inefficiency terms, \citet{barrios2021} introduced a logistic function; following this formulation, the inefficiency of firms is represented in Equation (2). Note that the high frequency $W_{t_j}$ and $R_{t_j}$ with several zero values are incorporated in Equation (2) and nonparametric functions.
\begin{equation}
u_{it}
=
\frac{1}
{1+\exp\!\left[
-\left(
\gamma_{1}Z_{1it}
+\gamma_{2}Z_{2it}
+\gamma_{3}Z_{3it}
+\sum_{j=1}^{12} f_{1}\!\left(R_{t_j}\right)
+\sum_{j=1}^{12} g_{1}\!\left(W_{t_j}\right)
\right)
\right]}
+\varepsilon_{it}.
\end{equation}

$f_j$'s and $g_j$'s are nonparametric functions of extreme weather events for month $j$ and $\varepsilon_{it} \sim N(0,\sigma_{\varepsilon}^{2})$, a random shock in the inefficiency component of the model. The effect of extreme weather events on inefficiency is relaxed through the nonparametric component in Equation (2) to adapt to the sparsity of the data on extreme weather events. Furthermore, the logistic function of inefficiency in Equation (2) will guarantee that $u_{it}$ from Equation (1) is non-negative. This can resolve the common challenges with truncated normal or half normal distributional assumption for $u_{it}$ as this has become a challenge for maximum likelihood estimation, see for example, \citet{kumbhakar2022}. The model presented in Equations (1) and (2) represents the varying frequency formulation of the stochastic frontier model (VF-SFM).

The parameters $\alpha_0$, $\alpha_1$, $\beta_1$, $\beta_2$, $\beta_3$, $\beta_4$, $\rho$, $\gamma_1$, $\gamma_2$, $\gamma_3$ and the nonparametric functions $f_j$ and $g_j$ in Equations (1) and (2) are estimated through a hybrid of ordinary least squares (OLS), generalized least squares (GLS), and penalized least squares (PLS) in a backfitting algorithm.

\subsection{Estimation Algorithm}

Equations (1) and (2) provide the necessary inputs in the analysis of stochastic efficiency among the firms and in the subsequent computation of their technical efficiencies. The iterative estimation process starts with initialization as follows:

\textit{Initialization:}

\begin{enumerate}
\item Ignoring $u_{it}$, estimate Equation (1) using GLS to obtain the initial estimates $\hat{\alpha}_0^{0}$, $\hat{\alpha}_1^{0}$, $\hat{\beta}_1^{0}$, $\hat{\beta}_2^{0}$, $\hat{\beta}_3^{0}$, $\hat{\beta}_4^{0}$ and $\hat{\rho}^{0}$.

\item Compute the residuals $e^0_{it}=\ln y_{it} -
 \ln \widehat{y}_{it}$, where $\ln \widehat{y}_{it}$ is based on the initial estimates $\hat{\alpha}_0^{0}$, $\hat{\alpha}_1^{0}$, $\hat{\beta}_1^{0}$, $\hat{\beta}_2^{0}$, $\hat{\beta}_3^{0}$, $\hat{\beta}_4^{0}$ and $\hat{\rho}^{0}$. Note that these residuals contain information on $-u_{it}$.

\item Estimate Equation (2) from   
\begin{equation}
u^*_{it} = -e^0_{it}
=
\frac{1}
{1+\exp\!\left[
-\left(
\sum^3_{j=1}\gamma_{j}Z_{jit}
+\sum_{j=1}^{12} f_{j}\!\left(R_{t_j}\right)
+\sum_{j=1}^{12} g_{j}\!\left(W_{t_j}\right)
\right)
\right]}
+\varepsilon_{it}.
\end{equation}

\begin{enumerate}[label=3.\arabic*.]
    \item Compute the logit of $u_{it}^*$ (and normalize to be between 0 and 1) to linearize Equation (3) while achieving the non-negativity of the inefficiency term, denote this by
    \begin{equation}
    m_{it} = \sum^3_{j=1}\gamma_{j}Z_{jit} +\sum_{j=1}^{12} f_{1}\!\left(R_{t_j}\right) +\sum_{j=1}^{12} g_{1}\!\left(W_{t_j}\right)
    \end{equation}
    \item Ignore $\sum_{j=1}^{12} f_{1}\!\left(R_{t_j}\right) +\sum_{j=1}^{12} g_{1}\!\left(W_{t_j}\right)$ from Equation (4) and estimate $\gamma_j, j = 1,2,3$, using OLS (resulting to $\widehat{\gamma}^0_{1}$,$\widehat{\gamma}^0_{2}$,$\widehat{\gamma}^0_{3}$) then compute new residuals by $m^0_{it} = m_{it} - \sum^3_{j=1}\widehat{\gamma}_j Z_{jit}$. These residuals contain information on $f_1,\ldots,f_{12}$ and $g_1,\ldots,g_{12}$. 
    \item Replace $m_it$ in Equation~(4) with $m_{it}^0$ to estimate $f_1$  and $g_1$ ignoring $f_2,\ldots,f_{12}$ and $g_2,\ldots,g_{12}$ via penalized least squares (e.g., splines as basis functions). The solutions $\widehat{f}_1$  and $\widehat{g}_1$ are splines. Update new residuals $m^{01}_{it} = m^0_{it} - \widehat{f}_1(R_{t_1}) - \widehat{g}_1(W_{t_1})$.

    \item Replace $m_{it}$ in Equation~(4) with $m^{01}_{it}$ to estimate $f_2$  and $g_2$ ignoring $f_3,\ldots,f_{12}$ and $g_3,\ldots,g_{12}$ via penalized least squares. Update new residuals $m^{02}_{it} = m^{01}_{it} - \widehat{f}_2(R_{t_2}) - \widehat{g}_2(W_{t_2})$.

    \item Iterate 3.3 and 3.4 until $\widehat{f}_{12}$ and $\widehat{g}_{12}$ are obtained.
    
\end{enumerate}

\item Predict $u_{it}$ from Equation (3) by plugging in the mean of estimates obtained from Steps 3.1 to 3.5 for each region in year $t$, say $\hat{u}_{it}^{0}$.

\item Initial estimates of parameters are $\hat{\alpha}_0^{0}$, $\hat{\alpha}_1^{0}$, $\hat{\beta}_1^{0}$, $\hat{\beta}_2^{0}$, $\hat{\beta}_3^{0}$, $\hat{\beta}_4^{0}$ and $\hat{\rho}^{0}$, the nonparametric functions from the iteration of Steps 3.3 and 3.4 are $\hat{f}_1^{0}, \hat{f}_2^{0}, \ldots, \hat{f}_{12}^{0}, \hat{g}_1^{0}, \hat{g}_2^{0}, \ldots, \hat{g}_{12}^{0}$ and the predicted value of inefficiency term is $\hat{u}_{it}^{0}$.

\end{enumerate}

After initial values of the parameters and the nonparametric functions are available, the iteration phase through the backfitting algorithm is implemented.

\textit{Iteration:} at the $r$-th iteration,

\begin{enumerate}
    \item 	Compute the residuals of the production function from Equation (1) with $\hat{u}_{it}^{r-1}$ predicted from $(r-1)$-th iteration. Use GLS to estimate the parameters of the production $\hat{\alpha}_0^{0}$, $\hat{\alpha}_1^{0}$, $\hat{\beta}_1^{0}$, $\hat{\beta}_2^{0}$, $\hat{\beta}_3^{0}$, $\hat{\beta}_4^{0}$ and $\hat{\rho}^{0}$.

\item Compute the residuals $e^r_{it}=\ln y_{it} -
 \ln \widehat{y}_{it}$, where $\ln \widehat{y}_{it}$ is based on the initial estimates $\hat{\alpha}_0^{r}$, $\hat{\alpha}_1^{r}$, $\hat{\beta}_1^{r}$, $\hat{\beta}_2^{r}$, $\hat{\beta}_3^{r}$, $\hat{\beta}_4^{r}$ and $\hat{\rho}^{r}$. Note that these residuals contain information on $-u_{it}$.

\item Estimate Equation (2) from   
\begin{equation}
u^*_{it} = -e^r_{it}
=
\frac{1}
{1+\exp\!\left[
-\left(
\sum^3_{j=1}\gamma_{j}Z_{jit}
+\sum_{j=1}^{12} f_{1}\!\left(R_{t_j}\right)
+\sum_{j=1}^{12} g_{1}\!\left(W_{t_j}\right)
\right)
\right]}
+\varepsilon_{it}.
\end{equation}

\begin{enumerate}[label=3.\arabic*.]
    \item Compute the logit of $u_{it}^*$ (and normalize to be between 0 and 1) to linearize Equation (3) while achieving the non-negativity of the inefficiency term, denote this by
    \begin{equation}
    m_{it} = \sum^3_{j=1}\gamma_{j}Z_{jit} +\sum_{j=1}^{12} f_{1}\!\left(R_{t_j}\right) +\sum_{j=1}^{12} g_{1}\!\left(W_{t_j}\right)
    \end{equation}
    
    \item Compute residuals from Equation (6), $m^{*r}_{it} = m^{r-1}_{it} - \sum_{j=1}^{12} \widehat{f}^{r-1}_{j}\!\left(R_{t_j}\right) - \sum_{j=1}^{12} \widehat{g}^{r-1}_{j}\!\left(W_{t_j}\right)$ and estimate $\gamma_j, j = 1,2,3$, using OLS (resulting to $\widehat{\gamma}^r_{1}$,$\widehat{\gamma}^r_{2}$,$\widehat{\gamma}^r_{3}$) then compute new residuals by $m^r_{it} = m^{*r}_{it} - \sum^3_{j=1}\widehat{\gamma}_j Z_{jit}$. These residuals contain information on $f_1,\ldots,f_{12}$ and $g_1,\ldots,g_{12}$. 
    
    \item Replace $m_it$ in Equation~(6) with $m_{it}^r$ adjusting for $\widehat{f}^{r-1}_2,\ldots,\widehat{f}^{r-1}_{12}$ and $\widehat{g}^{r-1}_2,\ldots,\widehat{g}^{r-1}_{12}$ to estimate $f_1$  and $g_1$ via penalized least squares. The solutions $\widehat{f}^{r}_1$  and $\widehat{g}^{r}_1$ are splines. Update new residuals $m^{r1}_{it} = m^r_{it} - \widehat{f}^{r}_1(R_{t_1}) - \widehat{g}^r_1(W_{t_1})$.

    \item Replace $m_it$ in Equation~(6) with $m_{it}^r$ adjusting for $\widehat{f}^{r-1}_1,\widehat{f}^{r-1}_3,\ldots,\widehat{f}^{r-1}_{12}$ and $\widehat{g}^{r-1}_1,\widehat{g}^{r-1}_3,\ldots,\widehat{g}^{r-1}_{12}$ to estimate $f_1$  and $g_1$ via penalized least squares. Update new residuals $m^{r2}_{it} = m^{r1}_{it} - \widehat{f}^{r}_2(R_{t_1}) - \widehat{g}^r_2(W_{t_1})$.

    \item Iterate 3.3 and 3.4 until $\widehat{f}^{r}_{12}$ and $\widehat{g}^{r}_{12}$ are obtained. After the iteration, $\widehat{\gamma}^r_1$,$\widehat{\gamma}^r_2$,$\widehat{\gamma}^r_3$,$\widehat{f}^r_1,\ldots,\widehat{f}^r_{12}$, $\widehat{g}^r_1,\ldots,\widehat{g}^r_{12}$ are obtained.
    
\end{enumerate}

\item Predict $u_{it}$ from Equation (5) by plugging in the mean of estimates obtained from Steps 3.1 to 3.5 for each region in year $t$, say $\hat{u}_{it}^{r}$.

\item Estimates of parameters of production function in this iteration are $\hat{\alpha}_0^{r}$, $\hat{\alpha}_1^{r}$, $\hat{\beta}_1^{r}$, $\hat{\beta}_2^{r}$, $\hat{\beta}_3^{r}$, $\hat{\beta}_4^{r}$ and $\hat{\rho}^{r}$, the parameters of inefficiency term are $\widehat{\gamma}^r_1$,$\widehat{\gamma}^r_2$,$\widehat{\gamma}^r_3$, and the estimates of the nonparametric functions are $\widehat{f}^r_1,\ldots,\widehat{f}^r_{12}$, $\widehat{g}^r_1,\ldots,\widehat{g}^r_{12}$. The predicted value of inefficiency term is $\hat{u}_{it}^{r}$.

\item 	Continue iteration from Step 1 until mean absolute prediction error (MAPE) convergence, i.e.,
\begin{equation*}
    \Delta\text{MAPE} = \text{MAPE}_r - \text{MAPE}_{r-1}< \tau ~~\text{(tolerance level)}.
\end{equation*}
		
Using estimates of parameters and nonparametric functions upon convergence, technical efficiency (TE) is then estimated from Equation (7) below.

\begin{equation}
\text{TE}_{it} = \exp \left\{
    \frac{-1}
{1+\exp\!\left[
-\left(
\sum^3_{j=1}\widehat{\gamma}_{j}Z_{jit} +\sum_{j=1}^{12} \widehat{f}_{j}\!\left(R_{t_j}\right) +\sum_{j=1}^{12} \widehat{g}_{j}\!\left(W_{t_j}\right)
\right)
\right]}
\right\}
\end{equation}

\end{enumerate}

\section{Results and Discussion}\label{chap:results}

The postulated VF-SFM model in Equations (1) and (2) are estimated from the spatio-temporal data on EC and compared to \citet{battese1995} with $W_{t_j}$ and $R_{t_j}$ aggregated to annual levels, i.e., maximum values for the year (Benchmark). Since \citet{battese1995} is estimated through MLE, models with different combination of capital indicators ($C$'s) and determinants of inefficiency ($Z$'s) were considered, but overall, three models dominated all others, these are:

\begin{itemize}
\item[\textbf{Model 1:}] Included all capital indicators ($C_1=\text{peak load}$; $C_2=\text{number of consumers}$; $C_3=\text{remittance to PSALM}$; $C_4=\text{reinvestment fund}$); and all determinants of inefficiency ($Z_1=\text{collection efficiency}$; $Z_2=\text{system loss}$; $Z_3=\text{interest expenses}$).

\item[\textbf{Model 2:}] Only $C_3=\text{remittance to PSALM}$ is included among capital indicators, and $Z_1=\text{collection efficiency}$ is included among the determinants of inefficiency.

\item[\textbf{Model 3:}] Only $C_3=\text{remittance to PSALM}$ is included among capital indicators, and $Z_2=\text{system loss}$ is included among the determinants of inefficiency.
\end{itemize}

The iterative estimation algorithm for VF-SFM discussed in Section 3.3 converged for all regions. On the other hand, estimation (MLE) of the benchmark model failed to converge for most regions, and in some regions, MLE was not even executed due to issues with residuals, see Table 3 for more details. Table 3 further shows that in cases where MLE for the benchmark model converged, VF-SFM is superior. In cases where MLE of the benchmark did not converge or that MLE is not feasible, VF-SFM still produced models that have reasonable predictive ability. It is clear that VF-SFM is able to capture the dynamics of production capability of firms when higher (than production indicators) frequency determinants of inefficiency are not aggregated to coincide with the frequency of production, labor, and capital indicators in spatio-temporal data. Disaggregated extreme weather events can rightfully capture stylized facts related to the threat this may cause for firm efficiency, especially for ECs.

\begin{table}[!t]
\centering
\caption{Mean Absolute Percentage Error of Different Models (GOR)}
\label{tab:mape_gor}
\begin{tabular}{ccccccc}
\hline
\multirow{2}{*}{\textbf{Region}} &
\multicolumn{2}{c}{\textbf{Model 1}} &
\multicolumn{2}{c}{\textbf{Model 2}} &
\multicolumn{2}{c}{\textbf{Model 3}} \\
\cline{2-7}
& \textbf{VF-SFM} & \textbf{Benchmark} & \textbf{VF-SFM} & \textbf{Benchmark} & \textbf{VF-SFM} & \textbf{Benchmark} \\
\hline
1  & 10.26 & 22,259.37 & 30.54 & 22,333.34 & 30.51 & 22,250.59 \\
2  & 13.63 & 16,313.93 & 100.00 & 16,700.82 & 100.00 & 16,511.79 \\
3  & 24.45 & 24,915.68 & 1142.05 & 25,202.89 & 1153.60 & * \\
4  & 21.24 & 26,867.71 & 100.00 & 28,416.00 & 100.00 & 28,692.01 \\
5  & 14.71 & 56,652.01 & 100.00 & 59,495.90 & 100.00 & 59,498.24 \\
6  & 20.89 & 14,322.73 & 100.00 & 14,497.08 & 100.00 & 14,373.35 \\
7  & 26.66 & 18,608.82 & 69.92 & 18,701.56 & 70.40 & 18,789.34 \\
8  & 21.10 & 24,626.24 & 41.68 & 22,557.56 & 14.27 & 22,577.25 \\
9  & 26.76 & * & 57.56 & 35,854.85 & 54.95 & 38,103.36 \\
10 & 35.35 & 12,488.28 & 58.66 & 10,561.65 & 58.36 & 9,620.31 \\
11 & 28.17 & 10,724.81 & 34.35 & 10,856.36 & 34.83 & 10,902.65 \\
12 & 21.67 & * & 65.50 & 14,185.35 & 66.34 & 14,127.28 \\
14 & 8.52  & 16,092.09 & 100.00 & 17,131.45 & 100.00 & 17,097.99 \\
15 & 100.00 & 40,162.14 & 100.00 & 39,647.40 & 100.00 & 40,462.79 \\
16 & 20.57 & 8,667.25 & 100.00 & 9,212.58 & 100.00 & 9,110.10 \\
17 & 33.46 & 9,984.97 & 100.00 & 10,197.47 & 100.00 & 10,828.37 \\
\hline
Overall & 25.72 &  & 208.72 &  & 210.07 &  \\
\hline
\end{tabular}

\vspace{0.2cm}
\begin{flushleft}
\footnotesize \textit{Note:} * MLE failed to converge.
\end{flushleft}
\end{table}

\subsection{Production and Inefficiency Functions (VF-SFM)}

Detailed characteristics of estimates from Model 1 are presented in this section. While Model 3 is parsimonious, Model 1 is advantageous compared to Models 2 and 3 in terms of capturing production and efficiency dynamics among the firms (EC). Similar insights are noted from the three models.

Estimates of the parameters of the production function by region are presented in Table 4. The backfitting algorithm, assuming additive models in Equations (1) and (2), was efficient in estimating parameters of the production function. Labor and Capital indicators are mostly significant over the regions. System Loss ($Z$), which competes with maximum rainfall ($R_{t_j}$) and maximum windspeed ($W_{t_j}$) in explaining inefficiency, are significant for most regions. However, whenever $Z$ is significant, the model correctly captures that increasing $Z$ (system loss) leads to more inefficiency ECs. The effects of Labor and Capital indicators are inconsistent across different regions, resulting from heterogeneity of ECs within a region, even after standardization of the output indicator GOR. Significance of remittance to PSALM as an indicator of capital is empirical evidence of the relevance of this regulatory agency in the power sector in the Philippines. This implies further that PSALM is serving its purpose of monitoring and supporting the ECs to ensure the viability of their operations through with a sound financial profile.

Autocorrelation of production indicator (GOR) is very prominent; this can be explained by the accumulation of technology (knowledge) by the firm that affects the production level. However, even with the growing amount of knowledge and technology, system loss confounded with vulnerability to extreme weather events can still pull down the efficiency of the ECs in distributing electricity in the countryside. See Table 5 for details.

\begin{table}[!t]
\centering
\caption{Parameter Estimates of Production Function for Model 1 (GOR)}
\label{tab:model1_prod}

\resizebox{\textwidth}{!}{%
\begin{tabular}{cccccc cccccc c}
\hline
\multirow{2}{*}{\textbf{Reg.}} &
\multicolumn{2}{c}{\textbf{Number of Employees}} &
\multicolumn{2}{c}{\textbf{Peak Load}} &
\multicolumn{2}{c}{\textbf{\# of Customers}} &
\multicolumn{2}{c}{\textbf{Remittance to PSALM}} &
\multicolumn{2}{c}{\textbf{Reinvestment Fund}} &
\multicolumn{2}{c}{$\hat{\rho}$} \\
\cline{2-13}
& \textbf{Est} &\textbf{ p-val} & \textbf{Est} & \textbf{p-val} & \textbf{Est} & \textbf{p-val} &
\textbf{Est} & \textbf{p-val} & \textbf{Est} & \textbf{p-val} & \textbf{Est} & \\
\hline
1  & .0816  & .2977 & .0172  & .4619 & 1.1302 & .0000 & .0360  & .1178 & .0706  & .0312 & .8806 & \\
2  & -.2276 & .0265 & .5916  & .0000 & .9366 & .0000 & -.0335 & .2417 & -.0514 & .0000 & .7363 & \\
3  & .0360  & .0132 & .1285  & .0228 & 1.1422 & .0000 & .0066  & .1719 & .0354  & .0732 & .9739 & \\
4  & -.0729 & .5837 & .5080  & .0005 & .9524 & .0000 & .0046  & .5113 & -.0962 & .2812 & .9336 & \\
5  & .0141  & .8467 & .6913  & .0000 & .5760 & .0000 & -.0049 & .6454 & .0821  & .0354 & .6659 & \\
6  & .0373  & .3518 & 1.0563 & .0000 & .1444 & .0000 & .1069  & .0000 & .0666  & .0421 & .9228 & \\
7  & -.0904 & .0895 & .0505  & .3117 & -.0187 & .3409 & .0404  & .0023 & 1.3627 & .0000 & .9209 & \\
8  & .6542  & .0000 & -.0590 & .0464 & -.0134 & .1516 & -.2434 & .0000 & 1.3605 & .0000 & .8568 & \\
9  & .3078  & .1342 & .0598  & .4988 & -.0253 & .4352 & .1297  & .0139 & 1.0808 & .0000 & .7549 & \\
10 & .4380  & .0002 & .4938  & .0000 & -.2501 & .0000 & .0159  & .4329 & .9019  & .0000 & .8581 & \\
11 & .2952  & .0792 & 1.0518 & .0000 & -.5133 & .0000 & .5267  & .0000 & .2303  & .0056 & .6316 & \\
12 & .2603  & .0138 & .7522  & .0000 & -.3716 & .0000 & .0539  & .0012 & .8446  & .0000 & .8306 & \\
14 & .0839  & .3017 & .3076  & .0000 & .3989 & .0000 & -.0366 & .1203 & .6593  & .0000 & .6570 & \\
15 & .0201  & .6625 & -.0092 & .2431 & .0097 & .1399 & .0157  & .0304 & .2416  & .0056 & .9999 & \\
16 & -.0735 & .4941 & .2237  & .0019 & -.0880 & .0011 & .2011  & .0000 & 1.1076 & .0000 & .8997 & \\
17 & -.0241 & .7852 & .2028  & .0423 & .0784 & .0319 & .0324  & .2291 & 1.0760 & .0000 & .9517 & \\
\hline
\end{tabular}
}
\end{table}

\begin{table}[!t]
\centering
\caption{Parameter Estimates of Inefficiency Function for Model 1 (GOR)}
\label{tab:model1_inefficiency}
\begin{tabular}{ccccccc}
\hline
\multirow{2}{*}{\textbf{Region}} &
\multicolumn{2}{c}{\textbf{Collection Efficiency}} &
\multicolumn{2}{c}{\textbf{Systems Loss}} &
\multicolumn{2}{c}{\textbf{Interest Expenses}} \\
\cline{2-7}
& \textbf{Estimate} & \textbf{p-value} & \textbf{Estimate} & \textbf{p-value} & \textbf{Estimate} & \textbf{p-value} \\
\hline
1  & .00798  & .0001 & -.00016 & .0589 & .00001 & .2724 \\
2  & -.02488 & .0000 & .20532  & .0000 & .00008 & .0001 \\
3  & .01324  & .0001 & -.08485 & .0102 & .00001 & .0104 \\
4  & -.04060 & .0001 & .03420  & .0000 & -.00001 & .4576 \\
5  & -.00695 & .0240 & .02897  & .1224 & .00002 & .0030 \\
6  & -.00372 & .5320 & -.00899 & .8653 & .00003 & .0520 \\
7  & .00468  & .0120 & .00001  & .0097 & .00002 & .4106 \\
8  & .00511  & .0009 & -.00001 & .0078 & .00002 & .1520 \\
9  & .01864  & .0002 & -.00000 & .5323 & .00000 & .7739 \\
10 & -.00776 & .0002 & -.00001 & .0014 & .00005 & .0000 \\
11 & .00778  & .1150 & -.00000 & .3680 & .00001 & .1946 \\
12 & .00580  & .3870 & -.00000 & .2763 & .00003 & .4005 \\
14 & .00151  & .8298 & -.00461 & .9256 & .00014 & .0000 \\
15 & .00341  & .5125 & -.00002 & .6491 & -.00000 & .2326 \\
16 & .00481  & .0110 & -.00000 & .1416 & .00002 & .0657 \\
17 & -.01710 & .0428 & .06868  & .3672 & .00013 & .0000 \\
\hline
\end{tabular}
\end{table}

\subsection{Aggregation of High Frequency Data}

After annual aggregation of maximum rainfall ($R_{t_j}$) and maximum windspeed ($W_{t_j}$), i.e., using maximum values for every region for the year, all indicators for the production function and the inefficiency equations are measured at annual frequency. The \citet{battese1995} model (benchmark), a fully parametric model is estimated with the same indicators as Models 1, 2, and 3, with the addition of maximum rainfall and windspeed as determinants of inefficiency in addition to $Z$.

The benchmark model is estimated through MLE with the inefficiency term assumed to follow a truncated normal distribution. Consistent with the observations in the literature, see, for example, \citet{papadopoulos2021}, model estimation generally suffers from convergence issues (see Table 3). Furthermore, from Table 3, the overall predictive ability of the benchmark model is inferior compared to the VF-SFM.

In regions where MLE converges, the importance of PSALM is again highlighted; parameter estimates are all positive except for Region 15. The signs of the number of employees (labor input), however are mostly negative, possibly due to heterogeneity of ECs within a region. See Table 6 for details.

\begin{table}[!t]
\centering
\caption{Estimates of Parameter of Production Function for Model 1 (GOR) (Regions with converging MLE)}
\label{tab:benchmark_prod}
\resizebox{\textwidth}{!}{%
\begin{tabular}{c cc cc cc cc cc}
\hline
\multirow{2}{*}{\textbf{Region}} &
\multicolumn{2}{c}{\makecell{\textbf{Number of Employees}\\(Labor)}} &
\multicolumn{2}{c}{\makecell{\textbf{Peak Load}\\\textbf{(Capital)}}} &
\multicolumn{2}{c}{\makecell{\textbf{No. of Customers}\\\textbf{(Capital)}}} &
\multicolumn{2}{c}{\makecell{\textbf{Remittance to PSALM}\\\textbf{(Capital)}}} &
\multicolumn{2}{c}{\makecell{\textbf{Reinvestment Fund}\\\textbf{(Capital)}}} \\
\cline{2-11}
& \textbf{Est} & \textbf{p-val} & \textbf{Est} & \textbf{p-val} & \textbf{Est} & \textbf{p-val} & \textbf{Est} & \textbf{p-val} & \textbf{Est} & \textbf{p-val} \\
\hline
1  & -0.413 & 0.022 & 0.137 & 0.000 & 0.839 & 0.000 & 0.038 & 0.218 & 0.333 & 0.000 \\
2  & 0.104 & 0.571 & 0.705 & 0.024 & 0.188 & 0.519 & 0.150 & 0.053 & -0.085 & 0.000 \\
3  & 0.056 & 0.956 & 0.687 & 0.492 & 0.124 & 0.901 & 0.071 & 0.943 & 0.071 & 0.943 \\
4  & -0.366 & 0.714 & 0.587 & 0.553 & 0.322 & 0.745 & 0.015 & 0.988 & 0.398 & 0.687 \\
5  & -0.072 & 0.769 & 0.675 & 0.019 & 0.296 & 0.095 & 0.009 & 0.591 & 0.234 & 0.119 \\
6  & -0.035 & 0.881 & 0.741 & 0.000 & 0.103 & 0.286 & 0.118 & 0.001 & 0.173 & 0.328 \\
7  & 0.023 & 0.981 & 0.165 & 0.869 & -0.069 & 0.945 & 0.078 & 0.938 & 0.843 & 0.399 \\
8  & 0.109 & 0.453 & 0.373 & 0.000 & 0.011 & 0.589 & 0.058 & 0.160 & 0.641 & 0.000 \\
9  & \multicolumn{10}{c}{*} \\
10 & -0.016 & 0.924 & 0.548 & 0.000 & 0.231 & 0.020 & 0.029 & 0.076 & 0.279 & 0.000 \\
11 & 0.208 & 0.836 & 0.773 & 0.440 & -0.383 & 0.702 & 0.331 & 0.741 & 0.074 & 0.941 \\
12 & \multicolumn{10}{c}{*} \\
14 & -0.136 & 0.076 & 0.668 & 0.000 & 0.019 & 0.867 & -0.009 & 0.796 & 0.271 & 0.031 \\
15 & -0.360 & 0.000 & 0.062 & 0.001 & 0.052 & 0.004 & -0.056 & 0.000 & 0.788 & 0.000 \\
16 & -0.093 & 0.926 & 0.321 & 0.748 & -0.114 & 0.909 & 0.120 & 0.904 & 0.931 & 0.352 \\
17 & 0.122 & 0.037 & 0.418 & 0.000 & -0.011 & 0.670 & 0.087 & 0.000 & 0.442 & 0.000 \\
\hline
\end{tabular}
}

\begin{flushleft}
\footnotesize \textit{Note:} * MLE failed to converge.
\end{flushleft}

\end{table}

In the inefficiency function, system loss is significant only in one region, and maximum sustained wind is significant in one region as well. For maximum rainfall, it is significant in one region, but the sign is negative, counterintuitive to the postulated vulnerability of efficiency of ECs from extreme weather events. 
In general, convergence issues, poor predictive ability, and inconsistent effect of determinants in the production and inefficiency function lead to the conclusion of inadequacy of the benchmark model to characterize the linkage between firm (EC) efficiency and extreme weather events in the distribution of electricity in the Philippine countryside. See Table 7 for details.

\begin{table}[!t]
\centering
\caption{Estimates of Parameter of Inefficiency Function for Model 1 (GOR) (Regions with Converging MLE) -- BC Model}
\label{tab:benchmark_inefficiency}
\resizebox{\textwidth}{!}{%
\begin{tabular}{c cc cc cc cc cc}
\hline
\multirow{2}{*}{\textbf{Region}} &
\multicolumn{2}{c}{\textbf{Collection Efficiency}} &
\multicolumn{2}{c}{\textbf{Systems Loss}} &
\multicolumn{2}{c}{\textbf{Investment Expenses}} &
\multicolumn{2}{c}{\textbf{Max Rainfall}} &
\multicolumn{2}{c}{\textbf{Maximum Sustained Wind}} \\
\cline{2-11}
& \textbf{Est} & \textbf{p-val} & \textbf{Est} & \textbf{p-val} & \textbf{Est} & \textbf{p-val} & \textbf{Est} & \textbf{p-val} & \textbf{Est} & \textbf{p-val} \\
\hline
1  & 0.005  & 0.781 & -0.000 & 0.988 & -0.000 & 0.761 & -0.001 & 0.720 & 0.001  & 0.509 \\
2  & -0.001 & 0.957 & -0.005 & 0.979 & -0.048 & 0.000 & -0.002 & 0.489 & -0.002 & 0.621 \\
3  & -0.002 & 0.998 & -0.000 & 0.999 & -0.094 & 0.912 & -0.027 & 0.978 & -0.021 & 0.984 \\
4  & -0.002 & 0.999 & -0.000 & 0.999 & -0.185 & 0.000 & -0.016 & 0.979 & -0.013 & 0.987 \\
5  & -0.037 & 0.945 & 0.032  & 0.970 & -0.015 & 0.000 & 0.007  & 0.936 & -0.012 & 0.786 \\
6  & -0.002 & 0.998 & -0.000 & 0.999 & -0.011 & 0.224 & 0.002  & 0.980 & -0.001 & 0.988 \\
7  & 0.002  & 0.999 & -0.097 & 0.000 & -0.117 & 0.000 & -0.061 & 0.952 & -0.565 & 0.572 \\
8  & 0.007  & 0.146 & -0.000 & 0.000 & 0.000  & 0.591 & -0.000 & 0.477 & -0.000 & 0.371 \\
9  & \multicolumn{10}{c}{*} \\
10 & 0.038  & 0.000 & -0.000 & 0.059 & -0.000 & 0.635 & 0.000  & 0.363 & -0.000 & 0.082 \\
11 & -0.000 & 0.999 & -8.604 & 0.000 & -4.116 & 0.000 & 0.002  & 0.999 & -0.013 & 0.989 \\
12 & \multicolumn{10}{c}{*} \\
14 & 0.012  & 0.000 & 0.035  & 0.000 & -0.000 & 0.000 & 0.000  & 0.069 & -0.000 & 0.000 \\
15 & 0.012  & 0.258 & -0.000 & 0.000 & 0.002  & 0.000 & -0.000 & 0.504 & -0.000 & 0.186 \\
16 & -0.000 & 0.999 & -0.083 & 0.000 & -5.619 & 0.000 & -0.000 & 0.983 & -0.066 & 0.947 \\
17 & 0.219  & 0.862 & 0.147  & 0.899 & -0.004 & 0.905 & -0.022 & 0.879 & -0.356 & 0.774 \\
\hline
\end{tabular}
}

\vspace{0.2cm}

\begin{flushleft}
\footnotesize \textit{Note:} * MLE failed to converge.
\end{flushleft}

\end{table}

\subsection{Estimates of Technical Efficiency}

Estimates of technical efficiency (TE) for the three models based on VF-SFM and the benchmark \citet{battese1995} are presented in Tables 8 and 9. The criticism of the benchmark model (and stochastic frontier models in general), including its estimation through MLE, has been featured in recent literature. For example, \citet{papadopoulos2024} noted that the positive skewness for inefficiency error leads to a singular information matrix, causing the failure of MLE to converge. Worst, technical efficiency estimates converging to 1 \citep{elmehdi2025, haschka2024}. This implies that the firms are producing at the frontier levels when they are not, and high estimates of technical efficiency result from misspecification of the inefficiency equation coupled with convergence issues in MLE. Estimates of TE for the VF-SFM and the benchmark model are summarized in Table 8 (by region) and Table 9 (by year).

The recent critique on SFM, i.e., wrong skewness leading to convergence of TE to 1, is clear from Tables 8 and 9. TE estimates from VF-SFM, however, are much lower than the TEs for the benchmark.

From the heatmap in Figure 1, low rainfall levels are observed in regions 1, 5, 6, 9, 10, 11, 12, and 15. TE in these regions is generally higher, indicating the generally efficient firms in these regions, as they are not vulnerable to inefficiency-inducing characteristics of an extreme amount of rainfall that can cause landslides and subsequent destruction of the electricity distribution lines.

\begin{table}[!t]
\centering
\caption{Estimates of Technical Efficiency by Region (GOR)}
\label{tab:te_region}
\begin{tabular}{ccccccc}
\hline
\multirow{2}{*}{\textbf{Region}} &
\multicolumn{2}{c}{\textbf{Model 1}} &
\multicolumn{2}{c}{\textbf{Model 2}} &
\multicolumn{2}{c}{\textbf{Model 3}} \\
\cline{2-7}
& \textbf{VF-SFM} & \textbf{Benchmark} & \textbf{VF-SFM} & \textbf{Benchmark} & \textbf{VF-SFM} & \textbf{Benchmark} \\
\hline
1  & 69.54 & 99.30 & 49.53 & 90.26 & 49.76 & 92.68 \\
2  & 61.57 & 100.00 & 39.30 & 73.51 & 35.57 & 84.90 \\
3  & 62.69 & 100.00 & 37.09 & 84.00 & 37.77 &  \\
4  & 50.83 & 100.00 & 52.74 & 43.54 & 52.99 & 37.17 \\
5  & 49.50 & 99.78 & 38.39 & 67.69 & 37.71 & 66.36 \\
6  & 44.51 & 99.97 & 49.36 & 85.08 & 49.63 & 95.77 \\
7  & 63.92 &  & 73.63 & 81.10 & 54.65 & 73.96 \\
8  & 57.46 & 32.97 & 64.54 & 78.75 & 52.82 & 78.05 \\
9  & 79.36 &  & 71.96 & 65.61 & 57.33 & 29.72 \\
10 & 46.97 & 13.36 & 51.19 & 28.27 & 46.05 & 99.99 \\
11 & 63.45 & 100.00 & 62.82 & 80.83 & 55.73 & 73.38 \\
12 & 67.15 &  & 49.75 & 43.39 & 47.92 & 36.11 \\
14 & 65.26 & 81.77 & 64.90 & 37.82 & 68.89 & 39.68 \\
15 & 48.80 & 87.06 & 48.78 & 82.38 & 47.66 & 84.11 \\
16 & 61.89 & 100.00 & 49.66 & 44.81 & 47.50 & 52.17 \\
17 & 45.26 & 99.96 & 42.77 & 75.45 & 41.89 & 33.82 \\
\hline
Overall & 57.35 & 79.40 & 51.48 & 69.13 & 47.32 & 68.48 \\
\hline
\end{tabular}
\end{table}

Some typhoons passing through PAR bring maximum sustained winds (MSW) of over 200 KPH. Furthermore, several of these typhoons may pass PAR several times a month, causing too much rainfall and distribution line posts falling down before restoration of electricity distribution by the ECs. Many of these typhoons occurred in 2022 and 2013. Several typhoons occurring in a month also happened in 2012, 2014, 2016, 2018, 2019, and 2020. The ECs during the years 2018 and 2019 also registered very high system losses, possibly brought about by strong winds from the typhoons during this period. In addition to high system losses in 2019, collection efficiency was also very low. The interaction of these factors can help explain the levels of TE, which is lowest in 2019 for Model 3 at 39\%. In 2022, while the typhoons brought very strong maximum sustained winds, because system losses were kept at low levels, TE is relatively high this year.

\begin{table}[!t]
\centering
\caption{Estimates of Technical Efficiency by Year (GOR)}
\label{tab:te_year}
\begin{tabular}{ccccccc}
\hline
\multirow{2}{*}{\textbf{Year}} &
\multicolumn{2}{c}{\textbf{Model 1}} &
\multicolumn{2}{c}{\textbf{Model 2}} &
\multicolumn{2}{c}{\textbf{Model 3}} \\
\cline{2-7}
& \textbf{VF-SFM} & \textbf{Benchmark} & \textbf{VF-SFM} & \textbf{Benchmark} & \textbf{VF-SFM} & \textbf{Benchmark} \\
\hline
2010 & 59.75 & 75.03 & 64.70 & 65.29 & 60.27 & 64.93 \\
2011 & 60.17 & 75.54 & 56.19 & 69.36 & 51.64 & 68.67 \\
2012 & 60.13 & 76.13 & 55.89 & 68.24 & 51.02 & 68.46 \\
2013 & 58.46 & 76.62 & 56.09 & 68.82 & 51.76 & 69.05 \\
2014 & 61.29 & 76.72 & 57.66 & 65.88 & 53.62 & 66.16 \\
2015 & 63.12 & 76.02 & 57.31 & 61.22 & 53.38 & 62.24 \\
2016 & 63.47 & 76.44 & 54.15 & 64.83 & 50.27 & 64.98 \\
2017 & 60.76 & 75.79 & 54.47 & 65.62 & 50.88 & 65.85 \\
2018 & 53.35 & 96.76 & 44.57 & 73.26 & 40.70 & 70.32 \\
2019 & 58.81 & 97.32 & 40.14 & 76.77 & 36.41 & 73.30 \\
2020 & 59.11 & 76.78 & 48.25 & 70.93 & 44.10 & 70.33 \\
2021 & 53.13 & 76.07 & 44.31 & 69.22 & 39.96 & 68.57 \\
2022 & 34.01 & 76.67 & 35.51 & 79.30 & 31.20 & 78.72 \\
\hline
Overall & 57.35 & 79.40 & 51.48 & 69.31 & 47.32 & 68.48 \\
\hline
\end{tabular}
\end{table}

\subsection{Effects of Extreme Weather Conditions on Firm Efficiency}

Reforms in the energy sector in the Philippines started in the 1990s. The goal was not only a stable and reasonably-priced in the service area, but also to expand coverage specially in the countryside. Two regulatory agencies were formed in the process, including the Energy Regulatory Commission (ERC) and the Power Sector Assets and Liabilities Management Corporation (PSALM). The agencies are mandated to provide technical assistance and capacitate cooperatives (rural enterprises) with financial management. While the electric cooperatives may be financially viable, extreme weather events are perpetual threats to efficiency of their production and operations. We have characterized the linkage between extreme weather events and efficiency of electric cooperatives. While extreme weather events cannot be controlled by these cooperatives, they can identify and adopt certain technologies that can mitigate the impact to the efficiency of their operation. In this case, the role of regulatory agencies cannot be undermined, they are crucial in the provision of technical knowledge to these cooperatives in the efficient distribution of electricity to its members.

\section{Empirical Experiments}\label{chap:experiments}

A simulation study was conducted to evaluate the performance of the proposed VF-SFM under varying cross-sectional $N$ (e.g., number of firms) and temporal dimension $T$ (length of time series data). The data-generating process comprised two groups of units (firms), with units within each group exposed to similar extreme weather conditions represented by high-frequency covariates in the inefficiency equation. One low-frequency input was generated for the production function, together with one low-frequency determinant of inefficiency and two high-frequency inefficiency determinants. To mimic mixed frequency settings, 12 high-frequency observations were generated for each low-frequency time period. Three panel data configurations were considered:  $N(24) < T(100)$; $N(24) = T(24)$, and; $N(100) > T(24)$. For each configuration, 200 Monte Carlo replications were performed, and model performance was assessed using the average technical efficiency (TE) estimates and the mean absolute percentage error (MAPE) across replications.

Model fit was controlled by aiming for specific values of the technical efficiency (TE) and the proportion of the value of the production function to the output variable. Low TE was generated around 0.38, Medium TE around 0.60, and High TE around 0.80. For the proportion of production function to the output, High is at 1.65 leading to around 43\% MAPE, Medium at 1.1 leading to a higher MAPE at 65\%, and Low at 1.05 leading to the highest MAPE at 87\%. Autocorrelations ($\rho$) simulated are 0.1, 0.5, and 0.9.

The true values of TE, estimated TE, and MAPE for the additive model and the benchmark \citep{battese1995} are presented by proportion of production function to the output in Table~\ref{tab:te_mape_prop} and by the true TE in Table~\ref{tab:te_mape_true_te}. The additive model estimates TE closer to the true TE values with increasing proportion of production function to output. The estimated TE from additive model is closer to the true value when $N=T$. The benchmark generally overestimates TE specially with higher proportion of production function to output. MAPE of the benchmark is about three times higher than that of the additive model and worst in case where $N<T$. By contrast, the additive model performs well when $N>T$. While the additive model also overestimates TE when the true TE is small, the benchmark model generally overestimates TE more severely.


\begin{table}[ht]
\centering
\caption{Estimated TE and MAPE by True Proportion of Production Function}
\label{tab:te_mape_prop}
\setlength{\tabcolsep}{2pt}
\renewcommand{\arraystretch}{0.8}


\begin{tabular}{ccccccc}
\hline
\multirow{2}{*}{\textbf{Sample Size}} &
\multirow{2}{*}{\renewcommand{\arraystretch}{0.5}\begin{tabular}{c}
     \textbf{True Prop.of}  \\
     \textbf{ Prod. Function}
\end{tabular}} &
\multirow{2}{*}{\textbf{True TE}} &
\multicolumn{2}{c}{\textbf{Estimated TE}} &
\multicolumn{2}{c}{\textbf{MAPE}} \\
\cline{4-7}
& & & \textbf{Additive} & \textbf{Benchmark} & \textbf{Additive} & \textbf{Benchmark} \\
\hline
\multirow{3}{*}{$N<T$}
& Low    & 0.698 & 0.497 & 0.665 & 93.510 & 306.621 \\
& Medium & 0.636 & 0.491 & 0.624 & 77.610 & 264.756 \\
& High   & 0.621 & 0.500 & 0.896 & 31.480 & 48.485 \\
\hline
\multirow{3}{*}{$N=T$}
& Low    & 0.748 & 0.538 & 0.855 & 84.725 & 269.192 \\
& Medium & 0.676 & 0.536 & 0.859 & 71.263 & 102.855 \\
& High   & 0.643 & 0.531 & 0.962 & 28.430 & 31.781 \\
\hline
\multirow{3}{*}{$N>T$}
& Low    & 0.703 & 0.456 & 0.796 & 92.981 & 79.512 \\
& Medium & 0.641 & 0.456 & 0.777 & 77.011 & 63.076 \\
& High   & 0.624 & 0.454 & 0.678 & 31.352 & 169.785 \\
\hline
\end{tabular}
\end{table}

\begin{table}[ht]
\centering
\caption{Estimated TE and MAPE by True TE}
\label{tab:te_mape_true_te}
\setlength{\tabcolsep}{2pt}
\renewcommand{\arraystretch}{0.8}

\begin{tabular}{ccccccc}
\hline
\multirow{2}{*}{\textbf{Sample Size}} &
\multirow{2}{*}{\renewcommand{\arraystretch}{0.5}\begin{tabular}{c}
     \textbf{True TE}  \\
     \textbf{Levels}
\end{tabular}} &
\multirow{2}{*}{\renewcommand{\arraystretch}{0.5}\begin{tabular}{c}
     \textbf{True TE}  \\
     \textbf{Values}
\end{tabular}} &
\multicolumn{2}{c}{\textbf{Estimated TE}} &
\multicolumn{2}{c}{\textbf{MAPE}} \\
\cline{4-7}
& & & \textbf{Additive} & \textbf{Benchmark} & \textbf{Additive} & \textbf{Benchmark} \\
\hline
\multirow{3}{*}{$N<T$}
& Low    & 0.377 & 0.502 & 0.885 & 49.520 & 60.429 \\
& Medium & 0.602 & 0.499 & 0.865 & 35.430 & 53.661 \\
& High   & 0.769 & 0.494 & 0.638 & 85.570 & 310.414 \\
\hline
\multirow{3}{*}{$N=T$}
& Low    & 0.380 & 0.524 & 0.861 & 44.437 & 78.506 \\
& Medium & 0.613 & 0.536 & 0.880 & 32.065 & 286.489 \\
& High   & 0.828 & 0.537 & 0.915 & 78.046 & 92.865 \\
\hline
\multirow{3}{*}{$N>T$}
& Low    & 0.378 & 0.447 & 0.796 & 46.788 & 79.512 \\
& Medium & 0.604 & 0.454 & 0.777 & 34.367 & 63.076 \\
& High   & 0.775 & 0.459 & 0.678 & 86.108 & 169.785 \\
\hline
\end{tabular}
\end{table}

\section{Conclusions}\label{chap:conclusion}

This study develops an alternative framework for assessing the impacts of climate change on firm-level sustainability by linking extreme weather events to production efficiency. Sustainability is viewed through the lens of productive efficiency, recognizing that firms operating closer to the production frontier are better positioned to utilize resources effectively while maintaining resilient and sustainable operations. Given that climate change increasingly manifests through more frequent and severe extreme weather events, understanding their effects on productive performance is critical for both managers and policymakers.

Methodologically, the study addresses two important limitations of conventional stochastic frontier models (SFMs). First, it overcomes the long-standing problem of “wrong skewness” and the associated convergence difficulties of maximum likelihood estimation by specifying inefficiency through a logistic function rather than imposing restrictive distributional assumptions on the inefficiency term. Second, it resolves the mismatch between low-frequency production data and high-frequency determinants of inefficiency by incorporating the latter through flexible nonparametric functions. This approach preserves valuable information that would otherwise be lost through temporal aggregation and allows for a richer characterization of the dynamic relationship between efficiency and external shocks. Estimation through a hybrid backfitting algorithm further provides a robust alternative to conventional MLE-based procedures.

The empirical application to electric cooperatives in the Philippines demonstrates that extreme weather events exert a significant influence on operational efficiency in electricity distribution. The results highlight the vulnerability of distribution utilities to climate-related disruptions and underscore the need for adaptive strategies that enhance resilience. Investments in modern technologies that reduce system losses, improve network reliability, and strengthen operational performance during adverse weather conditions can play a crucial role in mitigating inefficiency. At the same time, regulatory institutions such as the Energy Regulatory Commission (ERC) and the Power Sector Assets and Liabilities Management Corporation (PSALM) remain essential in supporting and monitoring electric cooperatives through capacity-building initiatives, sound financial management, technology adoption, and operational reforms aimed at improving collection efficiency and service delivery.

Overall, the proposed mixed-frequency stochastic frontier framework extends the methodological toolkit available for efficiency analysis and provides a practical approach for examining micro-level sustainability challenges arising from climate change. By capturing the effects of high-frequency environmental shocks on firm performance without sacrificing information, the model offers valuable insights for designing policies and strategies that promote efficiency, resilience, and sustainability in climate-vulnerable sectors.

\noindent%
\textbf{Acknowledgement:} The authors acknowledge the National Electrification Administration (NEA) and the Philippine Atmospheric, Geophysical and Astronomical Services Administration (PAG-ASA) for providing the data used in this study.

\noindent
\textbf{Codes:} Codes will be made available at GitHub when needed.

\noindent%
\textbf{Data Availability:} Anonymized data will be made available at GitHub.

\noindent%
\textbf{Competing Interest:} The authors have no competing interest nor potential conflict of interest to declare.

\noindent%
\textbf{Compliance with Ethical Standards:} The research did not involve human participants nor animals.















\bibliography{7_references}

\end{document}